# Industry 4.0:
## Challenges and Success Factors for Adopting Digital Technologies in Airports

Jia Hao Tan and Tariq Masood

University of Cambridge, Changi Airport Group, and University of Strathclyde

## Pre-Print

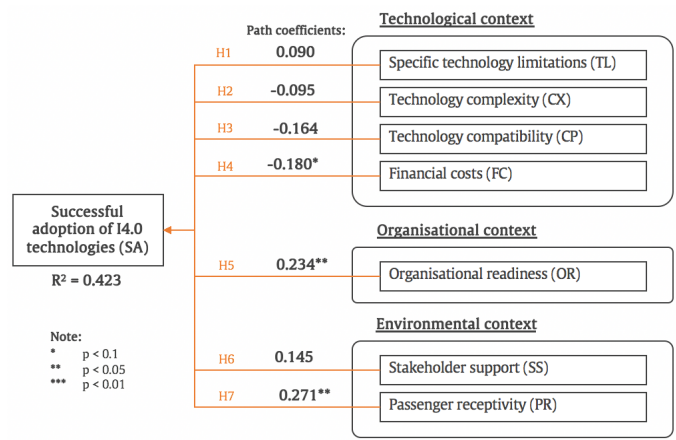

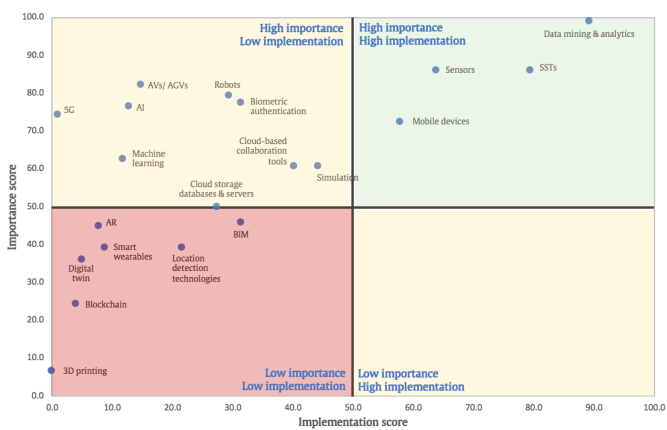

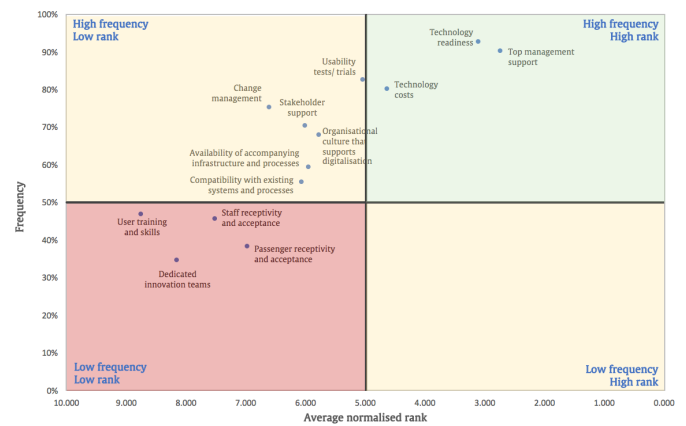

## Highlights

o   Airports have been undergoing digitalisation to capitalise on the purported benefits of Industry 4.0 technologies such as improved operational efficiency and passenger experience.

o   The COVID-19 pandemic has exacerbated the need for airports to adopt new technologies such as contactless and robotic technologies to facilitate travel during this pandemic. However, there is limited knowledge of recent challenges and success factors for adoption of digital technologies in airports.

o   Through an industry survey of airport operators and managers around the world (n=102, 0.754<Composite Reliability<0.892; conducted during COVID-19), this first-of-its-kind study identifies the implementation challenges faced in adopting Industry 4.0 technologies (n=20) as well as enhances understanding of best practices or success factors that supported technology adoption in airports.

o   The technology, organisation, environment (TOE) framework is used as a theoretically basis for the quantitative part of the questionnaire while a complementary qualitative part is used to underpin and extend the findings.

o   The results have shown that that the Industry 4.0 technologies were not implemented to a similar extent in airports despite the generic challenges that were faced in adopting the various Industry 4.0 technologies in the airport.

# Title: Industry 4.0: Challenges and success factors for adopting digital technologies in airports

**Authors:** Jia Hao Tan[1,2]*†, Tariq Masood[1,3]*†.

**Affiliations:**

[1]Institute for Manufacturing, Department of Engineering, University of Cambridge, 17 Charles Baggage Road, Cambridge, CB3 0FS, UK.

[2]Changi Airport Group (Singapore) Pte Ltd, 60 Airport Boulevard, Changi Airport Terminal 2, Singapore 819643.

[3]Department of Design, Manufacturing and Engineering Management, University of Strathclyde, 75 Montrose Street, Glasgow, G1 1XJ.

*Correspondence to: jht42@cantab.ac.uk (Jia Hao Tan); tariq.masood@strath.ac.uk (Tariq Masood)

†Equal contributions

**Abstract:**

With the advent of Industry 4.0 technologies in the last decade, airports have undergone digitalisation to capitalise on the purported benefits of these technologies such as improved operational efficiency and passenger experience. The ongoing COVID-19 pandemic with emergence of its variants (e.g. Delta, Omicron) has exacerbated the need for airports to adopt new technologies such as contactless and robotic technologies to facilitate travel during this pandemic. However, there is limited knowledge of recent challenges and success factors for adoption of digital technologies in airports. Therefore, through an industry survey of airport operators and managers around the world (n=102, 0.754<Composite Reliability<0.892; conducted during COVID-19), this study identifies the challenges faced in adopting Industry 4.0 technologies (n=20) as well as enhances understanding of best practices or success factors that supported technology adoption in airports. The widely used technology, organisation, environment (TOE) framework is used as a theoretically basis for the quantitative part of the questionnaire. A complementary qualitative part is used to underpin and extend the findings. The industry survey is the first-of-its-kind that was conducted to understand the implementation challenges that airport operators face in adopting Industry 4.0 technologies in the airport. The survey results have shown that that the Industry 4.0 technologies were not implemented to a similar extent in airports despite the generic challenges that were faced in adopting the various Industry 4.0 technologies in the airport.

**One Sentence Summary:** An industry survey on the challenges and success factors to develop airports of the future with Industry 4.0 technologies

# 1. Introduction

The airport industry has constantly faced a multitude of challenges. With the rise of the middle class and rapid economic development in Southeast Asian nations, the Asia-Pacific region is expected to drive the demand for air travel in the near future. Airport operators have put in place expansion plans to increase airport capacity to cater to burgeoning demand.

Within the first few months of 2020, the COVID-19 pandemic threw a curveball for the airport industry by greatly decimating air travel and presenting a fresh set of challenges for airport operators. From a state of near full capacity, airports now face historically low passenger numbers and have to step up on safe distancing measures and sanitisation operations.

Based on a recent structured literature review, airport operators have adopted a myriad of Industry 4.0 (I4.0) technologies over the years to deal with long-term capacity and efficiency challenges and also near-term challenges relating to the COVID-19 pandemic. While these technologies show demonstrated benefits for the airport industry, the adoption of these I4.0 technologies come with several technological, organisational and environmental challenges. These challenges, however, were only briefly discussed in literature with little inputs from airport operators. The objectives for adopting these I4.0 technologies by airport operators were also not clearly defined (Tan & Masood, 2021).

Given the lack of empirical research in the field of technology adoption in airports, an industry survey was adopted for empirical data collection. The survey serves as the primary source to collect both quantitative and qualitative inputs for the research study. As there is a lack of clarity on the important I4.0 technologies for future airports, the industrial survey will identify I4.0 technologies that are important for future airports along with the key objectives for digital transformation in airports. Furthermore, the survey also collects empirical inputs from airport operators on the challenges and key success factors for adopting I4.0 technologies in airports. The data obtained can be used for testing of the hypotheses relating to their implementation challenges.

Based upon the empirical data obtained from the surveys, a PLS-SEM statistical analysis was conducted on the quantitative data using the SmartPLS 3[1] software. The PLS-SEM analysis was adopted by similar technology adoption studies that involved latent variables relating to the TOE framework (Jia et al., 2017; Masood & Egger, 2019).

The rest of the article is structured as follows. Theoretical model is discussed in Section 2 while section 3 provides research hypotheses. Data collection is discussed in Section 4. Quantitative and quantitative results are presented in sections 5 and 6 respectively. Current state of adoption of I4.0 technologies in airports is discussed in section 7 before concluding the article in section 8.

# 2. Theoretical model

A literature review was first conducted to understand established technology adoption models that can potentially be used for the study (**Table 1**). This is followed by the development of a theoretical model to understand the challenges for adopting I4.0 technologies in airport.

---

[1] SmartPLS 3: https://www.smartpls.com

Table 1. Technology adoption models in literature.

| Model/ theory | Source | Key aspects/ constructs |
|---|---|---|
| Technology-Organisation-Environment (TOE) framework | Tornatsky and Fleischer (1990) | Technological innovation decision-making based on technology, organisation and (external task) environment. |
| Diffusion of Innovation (DOI) theory | Rogers (1995) | Organisational innovativeness based on individual (leader) characteristics, internal and external characteristics of organisation (organisational structure, system openness). |
| Institutional theory | Scott and Christensen (1995), Scott (2001) | Institutional environment (including social and cultural factors, concerns for legitimacy) are crucial in shaping organisational structure and decisions. |
| Iacovou et al. model | Iacovou et al. (1995) | Adoption of innovation based on perceived benefits, organisational readiness and external pressure. |
| Technology acceptance model (TAM) | Davis (1989) | Adoption of technology by individuals are based on perceived usefulness and perceived ease of use. |
| Theory of planned behaviour (TPB) | Ajzen (1985), Ajzen (1991) | An individual's behavioural intentions are shaped by attitude, subjective norms and perceived behavioural control. |
| Unified Theory of Acceptance and Use of Technology (UTAUT) | Venkatesh et al. (2003) | User intentions are based on performance expectancy, effort expectancy, social influence and facilitating conditions. |

As the research focuses on the adoption of I4.0 technologies by airport operators, technology adoption models that focus on technology adoption at the individual level such as the Technology Acceptance Model (TAM) (Davis et al., 1989), Theory of Planned Behaviour (TPB) (Ajzen, 1991) and the Unified Theory of Acceptance and Use of Technology (UTAUT) (Venkatesh et al., 2003) models are not relevant.

The Technology-Organisation-Environment (TOE) framework (Tornatzky & Fleischer, 1990), Diffusion of Innovation (DOI) theory (Rogers, 1995), institutional theory and the Iacovou et al. model (Iacovou et al., 1995) which focus on technology adoption at the firm level are most relevant. Out of these models, the TOE framework and DOI theory are most prevalently referenced in academic literature that studies technology adoption in firms (Oliveira & Martins, 2011). The TOE framework is preferred for this study as it covers the external task environment of the firm which is absent in the DOI theory. The environmental aspect is important for the study of airports since the airport environment comprises multiple stakeholders such as passengers, airlines and ground handling agencies. Based upon the TOE framework, the theoretical model for this research is developed and illustrated in **Fig 1**. It describes the impact of three antecedent areas (technological, organizational and environmental contexts) on the successful adoption of I4.0 technologies in airports.

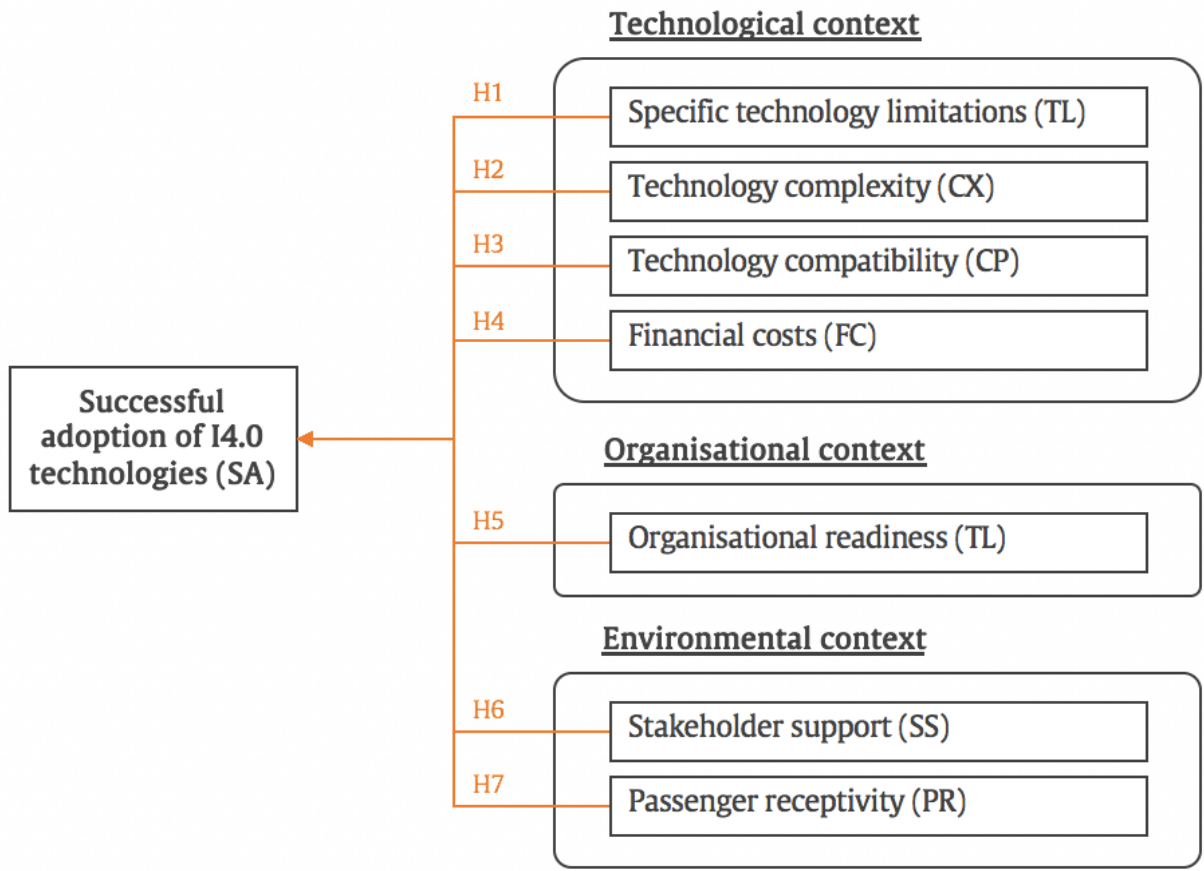

**Fig. 1. Theoretical model for research based upon TOE framework.**

## 3. Research hypotheses

This section describes the seven hypotheses that make up the theoretical framework presented in Fig. 1. These hypotheses were developed based on the challenges derived from the SLR (Tan & Masood, 2021). The dependent variable in the research is the successful adoption of I4.0 technologies in the airport environment (SA). For the purpose of this research, successful adoption refers to the organisation's willingness to increase the adoption of I4.0 technologies in the airport. This includes successful trials of I4.0 technologies in the airport environment which may lead to subsequent adoption on a wider scale in the airport within the next one to three years. The organisation also needs to obtain substantial benefits from adopting I4.0 technologies in the airport.

### *3.1 Specific technology limitations (TL)*

I4.0 technologies present a variety of technology-specific limitations within the airport environment depending on the type of technology and application involved. For example, the use of advanced data analytics for facial recognition in automated border control gates still require human security personnel as additional level of check as the technology is still subjected to errors (Opitz & Kriechbaum-Zabini, 2015; Sterchi & Schwaninger, 2015; Zhang & Jeong, 2017). Robots used for customer service still lack the ability to understand emotions unlike their human counterparts (Yau & Tang, 2018). The feasibility of robots and other autonomous vehicles is also limited by their battery lives, power consumption and the safety hazards they may potentially pose to other moving subjects or vehicles (Masson et al. 2017). These technology limitations reduce the relative advantage that new technologies have over the status quo, which may deter organisations from adopting new Industry 4.0 technologies if the benefits do not outweigh the costs (Tsai et al., 2010; Arnold et al., 2018). Hence, the following hypothesis is proposed:

**Hypothesis 1 (H1):** Specific technology limitations are **negatively** associated with the successful adoption of I4.0 technologies in airports.

## 3.2 Technology complexity (CX)

Technology complexity refers to the extent in which technologies are perceived to be challenging to learn, understand and adopt (Premkumar & Roberts, 1999). Given the complexity of new technology adoption, extensive usability trials will need to be carried out during the developmental phase to ensure usability and acceptance by end-users, and mitigate the risks of adoption (Oostveen et al. 2014). Furthermore, in order for I4.0 technologies to be successfully implemented in the airport environment, accompanying infrastructure, technologies and processes need to be in place. For example, a reliable Wi-Fi infrastructure is necessary to facilitate the use of mobile devices in the airport to complement the use of RFID luggage tagging and alert (Fei et al., 2016; Bogicevic et al., 2017; Shehieb et al., 2017). A network of sensors will also need to be deployed to facilitate real-time data monitoring and collection in the airport (Leung et al., 2013). The use of I4.0 technologies will also involve greater levels and frequency of data transfer and assimilation, especially for the purpose of data analytics and processing. This added level of complexity brings about greater threats of privacy leakages. This makes it challenging for organisations to adopt I4.0 technologies while having to preserve personal privacy, especially in an environment where regulations and existing policies for privacy protection are not mature enough (Onik et al., 2019). Hence, the following hypothesis is proposed:

**Hypothesis 2 (H2):** Technology complexity is **negatively** associated with the successful adoption of I4.0 technologies in airports.

## 3.3 Technology compatibility (CP)

Technology compatibility refers to the extent in which technological innovation can be easily integrated with existing infrastructure and processes (Prause, 2019). Companies are more inclined towards adopting new technologies if they fit existing systems and processes (Arnold et al., 2018). Given that airports vary in terms of design and operational aspects, the implementation of I4.0 technologies need to be customized to the specific airport context in order to meet specific operational requirements. For example, the design of automated people movers (APMs) lines need to accommodate specific terminal configurations of the airport (Little & Ross, 2013). Furthermore, established systems and processes within the airport such as the prevalent use of non e-passports may hinder the adoption of biometric and other advanced security technologies in the airport (Oostveen et al., 2014) Hence, the following hypothesis is proposed:

**Hypothesis 3 (H3):** Technology compatibility with existing infrastructure and processes is **positively** associated with the successful adoption of I4.0 technologies in airports.

## 3.4 Financial costs (FC)

In the context of new technology adoption, costs are not only incurred in the actual implementation of the technology, but are also significant in conducting usability trials, training of staff to implement new technologies, building of supporting IT infrastructure, maintenance, etc. While insufficient investments may prevent airports from reaping the full benefits of digital transformation, It is difficult to develop a business case for the benefits of new technology adoption to justify the costs incurred without detailed estimations (Premkumar & Roberts, 1999). Tornatzky and Klein (1982) have also highlighted that technologies that are perceived to be lower in costs are more likely to be adopted. Hence, the following hypothesis is proposed:

**Hypothesis 4 (H4):** Financial costs is **negatively** associated with the adoption of I4.0 technologies in airports.

## 3.5 Organisational readiness (OR)

Organisational readiness to change and embrace new technologies is essential for the successful adoption of I4.0 technologies (Jayashree et al., 2019). Organisational readiness can be analysed across several dimensions: (1) Organisational culture that promotes digitalization; (2) Top management involvement and support towards new technology adoption; (3) Digitalisation strategy/ roadmap for implementing Industry 4.0 technologies; (4) Staff receptivity towards new technology; (5) Relevant IT expertise among staff; (6) Human resource deployment towards innovation projects. (Tsai et al., 2010; Schüller & Wrobel, 2017; Jayashree et al., 2019; Sony & Naik, 2019) Ensuring organisational readiness across the various dimensions may pose a challenge for airports when adopting I4.0 technologies in airports. Hence, the following hypothesis is proposed:

**Hypothesis 5 (H5):** Organisational readiness is **positively** associated with the successful adoption of I4.0 technologies in airports.

## 3.6 Stakeholder support (SS)

Stakeholder support from external resources of an organisation contributes to the successful adoption of new technologies by ensuring that the technology solution is feasible, efficient and desirable for end-users (Prause, 2019). Airports need to engage other partners and stakeholders in the airport ecosystem (airlines, airport operators, regulatory agencies, software vendors and partners) to benefit from technological innovation (Stiffel, 2017). Technology vendors and partners also need to have technology solutions and systems ready to support the adoption of I4.0 technologies in the airport (Zhu et al., 2006). Hence, the following hypothesis is proposed:

**Hypothesis 6 (H6):** Stakeholder support is **positively** associated with the successful adoption of industry 4.0 technologies in airports.

## 3.7 Passenger receptivity (PR)

User receptivity or acceptance refers to the willingness and likelihood of users to adopt a new technology or system upon implementation (Fei et al., 2016). In the airport environment, passengers are the target end-users of most technologies, such as service robots for wayfinding, self-service kiosks for check-in and RFID luggage tags (Fei et al., 2016; Triebel et al., 2016; Singh et al., 2017). Oostveen et al. (2014) highlighted that involving airport end-users in the development of technology applications can help to better cater for the user requirements, thereby improving their receptivity and adoption of new technologies. Hence, the following hypothesis is proposed:

**Hypothesis 7 (H7):** Passenger receptivity is **positively** associated with the successful adoption of industry 4.0 technologies in airports.

## 4. Data collection

### 4.1 Survey questionnaire design

An 18-question survey questionnaire was developed using the Qualtrics XM[2] software to collect data relating to the research. The full questionnaire can be assessed in the link provided in the footnote[3]. The majority of the questions posed were closed-ended questions (Likert-scale, Yes/No, clearly defined options) to allow for easy classification of responses. Leading questions were omitted from the questionnaire to prevent introducing bias into the data obtained.

---
[2] Qualtrics XM software: https://www.qualtrics.com
[3] Survey questionnaire: https://cambridge.eu.qualtrics.com/jfe/form/SV_a4s6z0NkOnmTaHb

The quantitative aspect of the survey comprises of measures for the constructs which were adapted from past technology adoption studies and literature reviews to ensure validity and appropriateness in understanding the challenges of adopting I4.0 technologies in airports. These measures were placed on a five-point Likert-type scale ranging from strongly agree to strongly disagree. These measures, along with their sources, are summarized in **Table 2**.

The qualitative aspect of the survey comprises of multi-select and open-ended questions to gather deeper insights into three key areas of the research topic, namely the important technologies for future airports, objectives for digital transformation and the success factors for technology adoption.

Participants were given a multi-select question with a list of constituent technologies for I4.0 to indicate which technology is important for future airports. Participants were also allowed to provide alternative technologies that were not provided in the list. Similar questions were included for participants to indicate the important objectives and success factors for technology adoption in airports from a pre-defined list, while providing the option to indicate new items. Based upon their selection, participants have to rank the level of importance of the success factor for successful implementation of I4.0 technologies in the airport.

Apart from questions relating to the research variables, the questionnaire also contained miscellaneous questions to obtain general information about the respondents (designation, size of airport that their organisation operates, etc). The participants were also asked to provide their email addresses at the end of the survey if they were interested in participating in the semi-structured interviews relating to the research topic.

The survey questionnaire was pilot tested with I4.0 researchers to identify any discrepancies within the questionnaire and areas for improvement. Based on their inputs, the questionnaire was refined before being executed. An example of a refinement was the inclusion of a brief background of I4.0 technologies at the start of the survey to provide an overview of the constituent technologies and their functions. This served to facilitate an understanding of the research topic to improve the reliability of the survey responses.

### Table 2. Survey constructs and measures

| TOE category | Construct | | Measures | Source(s) |
|---|---|---|---|---|
| Technology | Specific technology limitations (TL) | TL1 | Industry 4.0 technologies have technology-specific limitations (e.g. efficiency, accuracy) that prevent it from effectively replacing existing technologies, systems or processes in airports. | Tsai et al. (2010), Arnold et al. (2018) |
| | | TL2 | Industry 4.0 technologies still require human supervision or human operators to meet operational requirements in airports. | |
| | | TL3 | The organisation is concerned about the health and safety hazards posed by the use of Industry 4.0 technologies within the airport environment. | |
| | Technology complexity (CX) | CX1 | Extensive usability tests or trials need to be carried out with end-users prior to implementing Industry 4.0 technologies in the airport. | Premkumar & Roberts (2010), Tsai et al. (2010) |
| | | CX2 | Appropriate IT infrastructure, technologies and processes need to be in place to complement the use of Industry 4.0 technologies in airports. | |
| | | CX3 | The organisation is concerned about cybersecurity and privacy concerns relating to the use of Industry 4.0 technologies within the airport environment. | |
| | Technology compatibility (CP) | CP1 | The implementation of Industry 4.0 technologies require few airport-specific adaptations/ customisations. | Arnold et al. (2018), Prause (2019) |
| | | CP2 | We can integrate the software necessary for Industry 4.0 technologies into our existing IT infrastructure with little effort. | |
| | Financial costs (FC) | FC1 | The costs of adopting Industry 4.0 technologies are far greater than the benefits. | Premkumar & Roberts (2010), Prause (2019) |
| | | FC2 | The costs for maintenance and support of Industry 4.0 technologies are very high for our business. | |
| | | FC3 | The amount of money and time invested in training staff to use Industry 4.0 technologies are very high. | |
| Organisation | Organisational readiness (OR) | OR1 | The management is supportive and actively develops a culture of digitalization within the organisation. | Premkumar & Roberts (2010), Tsai et al. (2010), Prause (2019) |
| | | OR2 | The organisation has set aside resources to train staff adopting Industry 4.0 technologies. | |
| | | OR3 | The organisation has dedicated teams that champion innovation or digitalization efforts in the airport. | |
| | | OR4 | The organisation has articulated a vision or strategy for digital transformation in the airport. | |
| | | OR5 | Staff is receptive towards the use of Industry 4.0 technologies in the airport. | |
| Environment | Stakeholder support (SS) | SS1 | Airport stakeholders were consulted before and during the implementation of Industry 4.0 technologies. | Zhu et al. (2006), Masood & Egger (2019) |
| | | SS2 | Technology vendors and partners have technology solutions and systems ready to support the adoption of Industry 4.0 technologies in the airport. | |
| | Passenger receptivity (PR) | PR1 | Passengers are actively involved in pilot tests before the implementation of Industry 4.0 technologies. | Zhu et al. (2010), Masood & Egger (2019) |
| | | PR2 | Passengers are receptive towards the use of Industry 4.0 technologies in the airport. | |

*4.2 Sampling frame*

The survey was sent out to a selection of 348 airport operators and managers around the world. These individuals were selected based upon two main criteria: (i) they are employees of organisations that operate or manage airports which are the target audience for the research study, and (ii) the individuals are involved in projects relating to I4.0 technologies in the course of their work. Thus, individuals who are involved in airport operations (e.g. baggage operations, passenger service), innovation and information technology (IT) were deemed to be suitable participants for the survey. These criteria were verified again in the questionnaire through a question on their job designation to ensure that the respondents have met the criteria.

348 prospective participants were shortlisted using the above criteria from professional social networks, speakers' lists from large airport-related trade shows (Passenger Terminal Expo) and industry contacts. An initial message was sent via email or online text message to introduce the research study and establish initial contact with the participant. Thereafter, for the participants who have indicated interest in the research study, an anonymous survey link was sent to the participant leading to the online questionnaire on Qualtrics.

Similar studies on technology adoption using the TOE framework were looked at to establish suitable sample sizes for the study. These studies had the following sample sizes: 38 (Prause, 2019), 65 (Zhu et al., 2010), 84 (Masood & Egger, 2019) and 206 (Jia et al., 2017). Based on a confidence level of 95% and a confidence interval of 0.1, the necessary sample size was calculated to be 97 using a sample size calculator[4]. After a six-week data collection process in June and July 2020, a total of 102 valid responses were obtained from the survey questionnaires, representing a response rate of 29.3%. The sample profiles for the survey are summarised in **Table 3**.

---

[4] Sample size calculator: https://www.abs.gov.au/websitedbs/D3310114.nsf/home/Sample+Size+Calculator

Tan, J.H. and Masood, T. (2021). "Industry 4.0: Challenges and success factors for adopting digital technologies in airports", pp. 1-25, pre-print, arXiv, 29/12/2021.
Page | 9

Table 3. Profile of survey respondents

| Sample size | | |
|---|---|---|
| Number of valid samples | 102 | |
| Response rate | 29.3% (102/348) | |
| | Frequency | Percentage (%) |
| **Designation** | | |
| C-suite (CEO, CIO, CTO, COO) | 5 | 4.9 |
| SVP, VP, AVP, Directors, Associate Directors, GM, AGM | 27 | 26.5 |
| Senior managers, managers, assistant managers | 59 | 57.8 |
| Senior associates, associates | 6 | 5.9 |
| Others | 5 | 4.9 |
| Total | 102 | 100.0 |
| **Organisation demographics** | | |
| **Type of organisation** | | |
| Government agency | 2 | 2.0 |
| Corporate entity | 95 | 93.1 |
| Others | 5 | 4.9 |
| Total | 102 | 100.0 |
| **Number of employees** | | |
| Below 10 | 0 | 0.0 |
| 11 - 250 | 5 | 4.9 |
| 251 - 1,000 | 40 | 39.2 |
| 1,001 - 10,000 | 55 | 53.9 |
| Above 10,000 | 2 | 2.0 |
| Total | 102 | 100.0 |
| **Number of passengers served by airport (2019)** | | |
| Less than 10 million | 7 | 6.9 |
| 10 - 30 million | 8 | 7.8 |
| 30 - 50 million | 29 | 28.4 |
| 50 - 70 million | 49 | 48.0 |
| More than 70 million | 9 | 8.8 |
| Total | 102 | 100.0 |

## 5. Quantitative results

The quantitative results from the survey provides a test of the hypotheses raised earlier about the challenges that airport operators face in adopting I4.0 technologies.

The quantitative results were first tested for validity and reliability before a partial-least square regression method (PLS-SEM) was used for statistical analysis to test the various hypotheses.

### *5.1 Convergent validity and reliability*

The measures were first assessed on whether they are good measurements for the latent variables. Measures which have a reflective indicator loading above 0.5 is deemed to be a good measurement of the latent construct (Hulland, 1999). The measure TL3 was removed from the model as its reflective indicator loading is below 0.5. The bolded values in **Table 4** show the outer loadings of the remaining measures for their associated latent variables. The composite reliability (CR) assesses the internal consistency of the measures in measuring their associated constructs. **Table 5** shows that the CR values are all greater than 0.7 which indicates adequate internal consistency of the constructs (Gefen et al., 2000). The convergent reliability of the constructs is assessed using the Average Variance Extracted (AVE) which indicates the proportion of variance that is explained by the latent variable. **Table 5** shows that the AVE for most of the latent variables are above the threshold value of 0.5 which is satisfactory (Bagozzi & Yi, 1988). Although the AVE for OR is 0.475, an AVE between 0.4 to 0.5 is still acceptable if the composite reliability is above 0.6 (Fornell & Larcker, 1981).

Table 4. Loadings and cross-loadings of measures onto latent variables

| Measures | Latent variable | | | | | | | |
|---|---|---|---|---|---|---|---|---|
| | Specific technology limitations (TL) | Technology complexity (CX) | Technology compatibility (CP) | Financial costs (FC) | Organisational readiness (OR) | Stakeholder support (SS) | Passenger receptivity (PR) | Successful adoption (SA) |
| TL1 | 0.872 | 0.127 | 0.229 | 0.104 | -0.142 | -0.147 | -0.178 | -0.174 |
| TL2 | 0.766 | 0.431 | 0.246 | 0.316 | -0.259 | -0.052 | -0.348 | -0.133 |
| CX1 | 0.314 | 0.830 | 0.071 | 0.323 | 0.051 | 0.006 | -0.185 | -0.183 |
| CX2 | 0.205 | 0.506 | -0.014 | 0.306 | 0.004 | 0.000 | -0.057 | -0.017 |
| CX3 | 0.196 | 0.822 | 0.351 | 0.228 | 0.035 | -0.006 | -0.082 | -0.178 |
| CP1 | 0.282 | 0.267 | 0.986 | 0.221 | -0.243 | -0.168 | -0.277 | -0.383 |
| CP2 | 0.167 | 0.020 | 0.596 | 0.113 | -0.173 | -0.116 | -0.092 | -0.079 |
| FC1 | 0.090 | 0.157 | 0.152 | 0.690 | -0.118 | 0.084 | -0.058 | -0.270 |
| FC2 | 0.188 | 0.250 | 0.098 | 0.778 | -0.046 | -0.186 | -0.366 | -0.269 |
| FC3 | 0.245 | 0.347 | 0.236 | 0.722 | -0.076 | -0.033 | -0.371 | -0.257 |
| OR1 | -0.076 | 0.136 | -0.099 | -0.070 | 0.645 | 0.284 | 0.170 | 0.221 |
| OR2 | -0.094 | -0.058 | -0.220 | -0.118 | 0.594 | 0.116 | 0.229 | 0.170 |
| OR3 | -0.184 | 0.115 | -0.109 | 0.068 | 0.722 | 0.208 | 0.254 | 0.279 |
| OR4 | -0.185 | -0.007 | -0.073 | -0.026 | 0.759 | 0.322 | 0.298 | 0.372 |
| OR5 | -0.214 | 0.001 | -0.368 | -0.233 | 0.714 | 0.353 | 0.282 | 0.345 |
| SS1 | -0.096 | 0.011 | -0.269 | -0.033 | 0.402 | 0.933 | 0.305 | 0.372 |
| SS2 | -0.128 | -0.024 | 0.135 | -0.093 | 0.157 | 0.601 | 0.375 | 0.168 |
| PR1 | -0.277 | -0.172 | -0.187 | -0.412 | 0.365 | 0.400 | 0.895 | 0.455 |
| PR2 | -0.269 | -0.117 | -0.290 | -0.236 | 0.291 | 0.308 | 0.899 | 0.464 |
| SA1 | -0.238 | -0.240 | -0.414 | -0.281 | 0.390 | 0.263 | 0.500 | 0.903 |
| SA2 | -0.152 | -0.185 | -0.300 | -0.391 | 0.436 | 0.402 | 0.451 | 0.913 |
| SA3 | 0.001 | -0.032 | -0.068 | -0.170 | 0.104 | 0.225 | 0.242 | 0.562 |

**Table 5. CR, AVE and VIF of respective latent variables**

| Latent variable | CR | AVE | VIF |
|---|---|---|---|
| Specific technology limitations (TL) | 0.805 | 0.674 | 1.266 |
| Technology complexity (CX) | 0.772 | 0.540 | 1.277 |
| Technology compatibility (CP) | 0.788 | 0.664 | 1.209 |
| Financial costs (FC) | 0.775 | 0.535 | 1.290 |
| Organisational readiness (OR) | 0.818 | 0.475 | 1.350 |
| Stakeholder support (SS) | 0.754 | 0.616 | 1.307 |
| Passenger receptivity (PR) | 0.892 | 0.805 | 1.503 |
| Successful adoption (SA) | 0.845 | 0.655 | |

## 5.2 Discriminant validity

Discriminant validity assesses whether each measure is subjectively independent from other measures on its latent variable (Mooi & Sarstedt, 2011). Three methods were used to assess the discriminant validity of the quantitative results, namely the cross-loading criterion, the Fornell and Larcker criterion and the Heterotrait-Monotrait (HTMT) ratio of correlations. The cross-loading criterion is satisfied as the measures load more strongly on their associated constructs as compared to their cross-loadings on other constructs (**Table 4**) (Chin, 1998). The Fornell and Larcker criterion is also satisfied as the square-root of the AVE of the latent variables are higher than the correlations between the latent variable and all other variables (Fornell & Larcker, 1981). **Table 6** shows the correlations between the various constructs with the diagonals representing the square-roots of the AVE. Furthermore, it is evident from **Table 7** that all the HTMT values fall below the threshold value of 0.9 which indicates good discriminant validity (Henseler et al., 2015).

**Table 6. Discriminant validity (Fornell and Larcker criterion)**

| Latent variable | CP | CX | FC | OR | PR | SA | SS | TL |
|---|---|---|---|---|---|---|---|---|
| CP | 0.815 | | | | | | | |
| CX | 0.245 | 0.735 | | | | | | |
| FC | 0.221 | 0.342 | 0.731 | | | | | |
| OR | -0.251 | 0.051 | -0.11 | 0.689 | | | | |
| PR | -0.266 | -0.161 | -0.36 | 0.365 | 0.897 | | | |
| SA | -0.36 | -0.213 | -0.364 | 0.424 | 0.512 | 0.809 | | |
| SS | -0.173 | 0 | -0.062 | 0.393 | 0.394 | 0.372 | 0.785 | |
| TL | 0.286 | 0.313 | 0.237 | -0.233 | -0.304 | -0.189 | -0.128 | 0.821 |

Table 7. Discriminant validity (HTMT)

| Latent variable | CP | CX | FC | OR | PR | SA | SS | TL |
|---|---|---|---|---|---|---|---|---|
| CP | | | | | | | | |
| CX | 0.306 | | | | | | | |
| FC | 0.347 | 0.623 | | | | | | |
| OR | 0.387 | 0.16 | 0.245 | | | | | |
| PR | 0.313 | 0.207 | 0.557 | 0.478 | | | | |
| SA | 0.356 | 0.259 | 0.54 | 0.514 | 0.659 | | | |
| SS | 0.454 | 0.204 | 0.337 | 0.592 | 0.75 | 0.593 | | |
| TL | 0.461 | 0.575 | 0.474 | 0.389 | 0.508 | 0.325 | 0.362 | |

*5.3 Hypothesis testing*

The PLS-SEM method is used to test the research hypotheses outlined earlier. Aside from the validity and reliability tests in the previous section, the quantitative results are further assessed for multi-collinearity among the latent variables by calculating the Variance Inflation Factor (VIF) using SmartPLS 3. **Table 5** shows that the VIF values for the latent variables are below the threshold level of 10 which suggests that multi-collinearity is not an issue with the dataset and the quantitative data is suitable for regression analysis (Mooi & Sarstedt, 2011).

A PLS-SEM model was constructed using the SmartPLS 3 software to capture the relationship among the latent variables with their associated measures. Next, the PLS-SEM algorithm was run to obtain the $R^2$ values and path coefficients of the latent variables. The $R^2$ value obtained for the independent variable is 0.423, which means that 42.3% of the variance in the dependent variable (SA) can be explained by the latent variables in the model. Path coefficients can take values between -1 and 1, where higher absolute values denote stronger predictive relationships of the latent variables. The path coefficients are shown on the arrows from the latent variables to the dependent variable in **Fig. 2**. As PLS-SEM does not assume a normal distribution in the sample, conducting bootstrapping was necessary to obtain t-statistics and path significance levels for each of the hypothesized relationship associated with the latent variables (Hair, 2014). **Fig. 2** and **Table 8** show the results from the bootstrapping involving 5,000 sub-samples using the SmartPLS 3 software. From the results, only Hypothesis 4, Hypothesis 5 and Hypothesis 7 are accepted as their respective latent variables (financial costs, organisational readiness and passenger receptivity) significantly influence the successful adoption of I4.0 technologies in the airports.

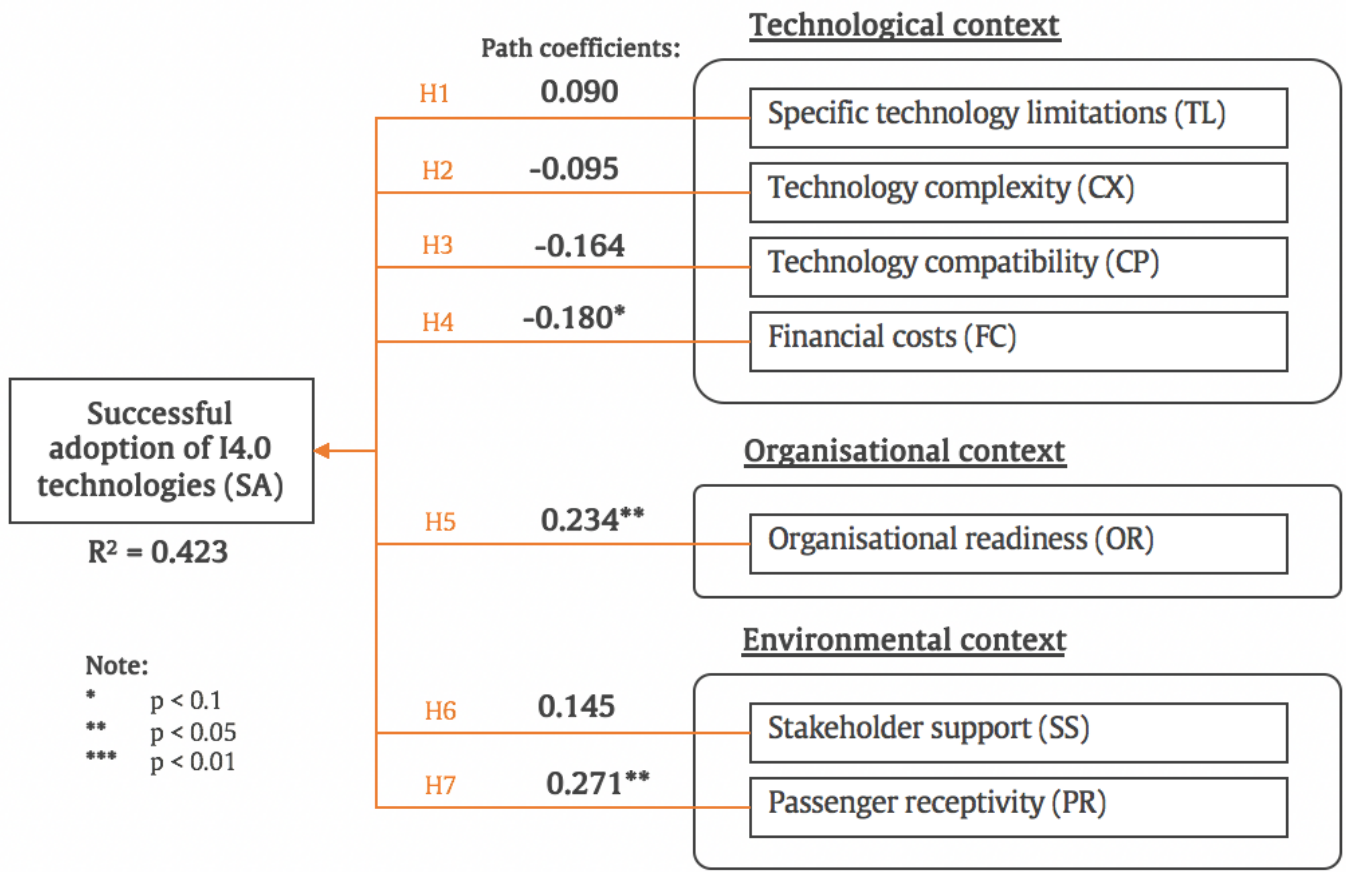

**Fig. 2. Theoretical model with empirical results.**

Table 8. Hypothesis testing results from PLS-SEM analysis

| Hypothesis | Relationship | Path coefficient | p-values | Significance | Decision |
|---|---|---|---|---|---|
| H1 | Specific technology limitations (TL) -> Successful adoption of I4.0 technologies (SA) | 0.090 | 0.457 | p > 0.1 | rejected |
| H2 | Technology complexity (CX) -> Successful adoption of I4.0 technologies (SA) | -0.095 | 0.398 | p > 0.1 | rejected |
| H3 | Technology compatibility (CP) -> Successful adoption of I4.0 technologies (SA) | -0.164 | 0.152 | p > 0.1 | rejected |
| H4 | Financial costs (FC) -> Successful adoption of I4.0 technologies (SA) | -0.180 | 0.07 | p < 0.1 | accepted |
| H5 | Organisational readiness (OR) -> Successful adoption of I4.0 technologies (SA) | 0.234 | 0.015 | p < 0.05 | accepted |
| H6 | Stakeholder support (SS) -> Successful adoption of I4.0 technologies (SA) | 0.145 | 0.141 | p > 0.1 | rejected |
| H7 | Passenger receptivity (PR) -> Successful adoption of I4.0 technologies (SA) | 0.271 | 0.012 | p < 0.05 | accepted |

## 6. Qualitative results

The qualitative results from the surveys present deeper insights into the important technologies for future airports, the objectives for digital transformation and the success factors for technology adoption.

### 6.1 Objectives for adopting I4.0 technologies in airports

In the survey questionnaire, airport operators were provided with a multi-select question with a pre-defined list of objectives for adopting I4.0 technologies based upon the initial literature review. The second column in Table 9 shows how often the pre-defined objectives were selected in the survey in percentage terms.

Airport operators were also presented with an open-ended question to provide other objectives for adopting I4.0 technologies aside from the pre-defined list of objectives. Based upon the responses from the open-ended survey questions, similar inputs were consolidated and 3 new objectives (reduce operational costs, overcome manpower shortages, mitigate health and safety concerns) were added to the list (Table 9). The relevant qualitative inputs from the surveys are extracted and presented in the rightmost column of the table.

### Table 9. Objectives for adopting I4.0 technologies in airports

| | Objectives for adopting I4.0 technologies in airports | Frequency | Qualitative inputs from surveys and interviews |
|---|---|---|---|
| **Pre-defined from literature review** | Improve operational efficiency | 100.0% | · Make for more targeted selection of passengers for screening and increase security screening accuracy and consistency (INT-5)<br>· Better monitoring and utilisation of assets within the airport.<br>· Make more data-driven decisions such as better matching demand and supply of ground assets (INT-6, INT-9) |
| | Enhance passenger experience | 93.1% | · Quick response to customer needs/ experience |
| | Boost capacity from existing infrastructure | 84.3% | · Future-proofing<br>· Future planning and life-cycle assessment |
| | Better respond to uncertain/ random events | 64.7% | · Future-proofing.<br>· Make use of digital technologies to leverage capabilities in an agile manner (INT-4) |
| | Generate value/ ancillary revenue | 50.0% | · New business model opportunity<br>· Provide greater on-demand transactional offerings on digital platforms (INT-4) |
| **Newly-added from data collected** | Reduce operational costs | New | · Cost reduction in the long run<br>· Reduce OPEX cost<br>· More cost effective solutions in the long term |
| | Overcome manpower shortages | New | · Manpower reduction in the long run through increased staff productivity<br>· Address manpower constraints and optimise resources<br>· Reduce over-reliance on manpower (INT-7, INT-8) |
| | Mitigate health and safety concerns | New | · Biosecurity and self-service social distancing<br>· Health concerns (pandemic management)<br>· Facilitate ongoing terminal sanitisation efforts with reduced manpower (INT-11) |

## 7. Current state of adoption of I4.0 technologies in airports

In the survey questionnaire, airport operators were provided with a list of I4.0 technologies that were identified from the initial literature review (Tan & Masood, 2021). From this list, airport operators had to identify the technologies which they perceive to be important for future airports based on their airport management experience. Thereafter, they were tasked to indicate the technologies which are currently implemented within the airports that they operate based upon their selections. Two scores were computed based on how often the technology was selected in the survey in percentage terms.

- **Implementation score:** Reflects how often the particular I4.0 technology is implemented in current airports.

- **Importance score:** Reflects the relative importance of the I4.0 technology for future airports.

The implementation scores were calculated based on how often survey participants have indicated that the I4.0 technology was implemented in airports currently. The importance score was calculated based on how often the survey participants have indicated that the technology was important for future airports. Fig. 3 provides a snapshot of the current state of adoption of I4.0 technologies in the airport based on their importance and implementation scores. The I4.0 technologies shown in the top-right quadrant (green) are considered to be both important and are widely implemented in airports currently.

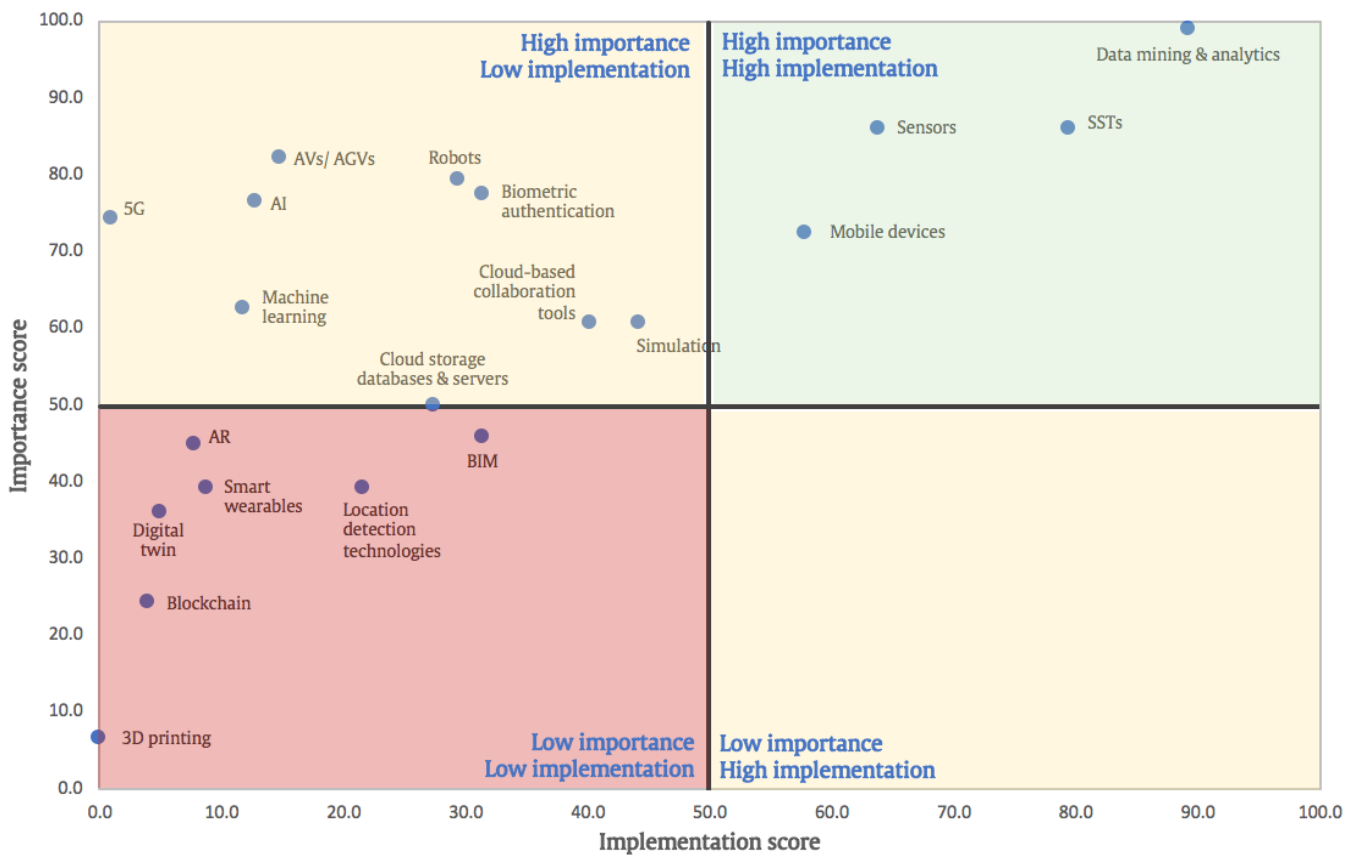

**Fig. 3. Implementation and importance scores of I4.0 technologies in airports.**

*7.1 Success factors for adopting I4.0 technologies in airports*

In the survey questionnaire, airport operators were also tasked to identify key success factors for successful adoption of I4.0 technologies in the airport from a pre-defined list of success factors based on literature review and interview inputs. For airport operators who have indicated that the airports they manage do not have any past successes in implementing I4.0 technologies, their responses in this category were excluded. Thereafter, they were tasked to rank the success factors based upon their selections. Two sets of data were generated from the data collected:

- **Frequency:** Percentage of how often the factor was selected by survey respondents to be a success factor for I4.0 technology adoption.

- **Average normalised rank:** Based on the ranking that survey respondents have indicated for their selection of success factors, normalisation was carried out to account for the different number of success factors that may have been selected by different respondents. Thereafter, the average of the normalised ranks for each success factor was computed. The normalised rank for each success factor can range from 1 to 10, with 1 being the most important success factor.

Fig. 4 illustrates the various success factors mapped according to their frequencies and average normalised rank in the survey. The success factors in the top-right quadrant (green) were most frequently selected by survey respondents and also ranked highly in terms of their importance for I4.0 technology adoption in airports.

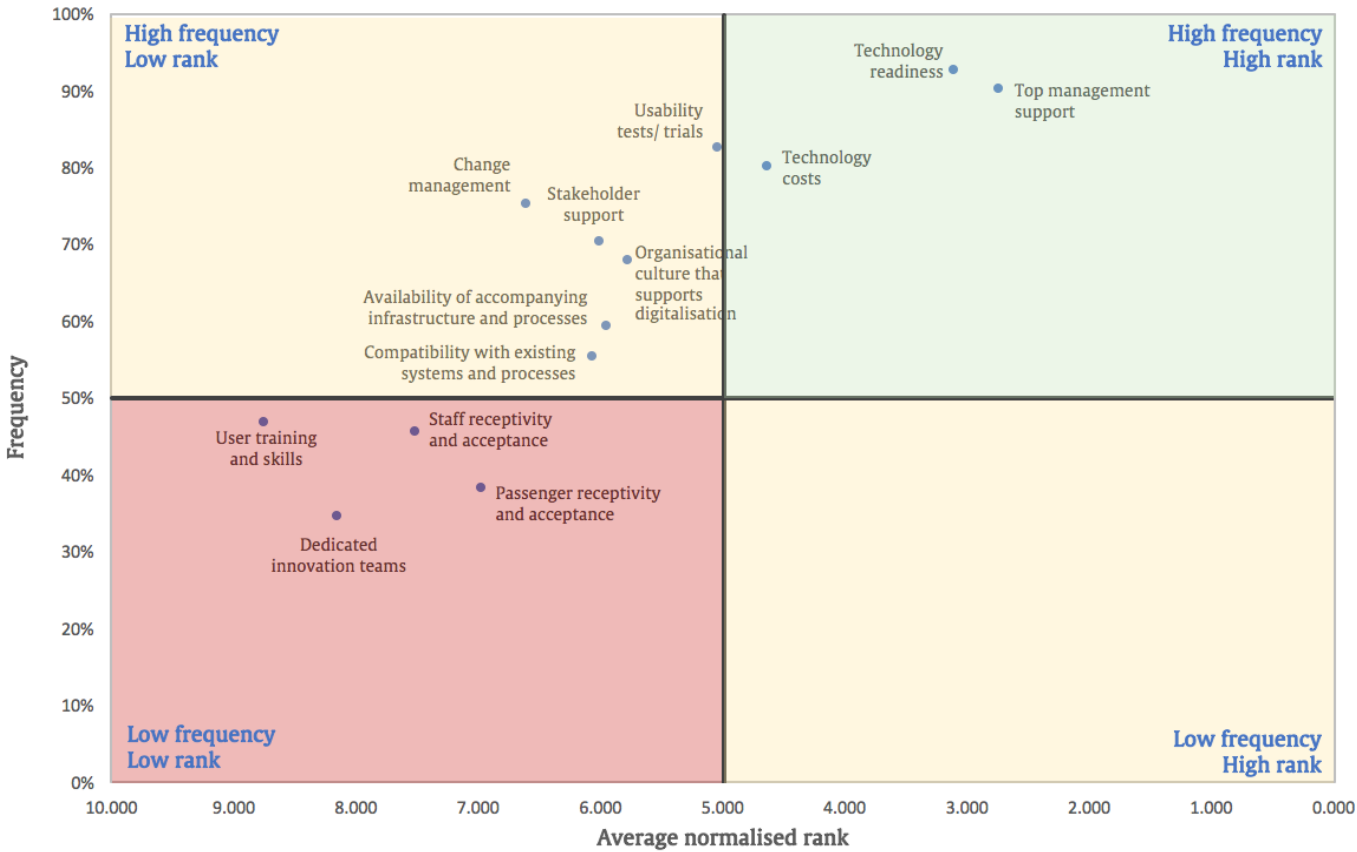

**Fig. 4. Frequencies and average normalised ranks of success factors**

## 8. Discussion

The results from the industry survey has surfaced three new objectives for adopting Industry 4.0 technologies in airports, namely to overcome manpower shortages, mitigate health and safety concerns as well as reduce operational costs. The nature of these new objectives can be attributed to the ongoing COVID-19 pandemic which has placed immense pressure on airports to create an environment for safe travel while overcoming manpower shortages and cost pressures. This reflects the increased demand by the airport industry for Industry 4.0 technologies due to its purported benefits.

The survey has also provided a perspective towards the type of Industry 4.0 technology that is most relevant for airports. According to the survey results (see **Fig. 3**), data mining and analytics is the most widely adopted technology along with sensors for IoT, CPS such as SSTs and mobile devices, which is consistent with the literature (Tan & Masood, 2021). Furthermore, the fairly prevalent implementation of cloud computing solutions such as cloud-based collaboration tools and cloud storage databases and servers was consistent with observations from Yau and Tang (2018) that cloud computing is a key enabler technology for other I4.0 technology applications even though it was underrepresented in the papers reviewed.

It is interesting to note the technologies which fall within the top-left quadrant (yellow) of Fig. 3. These technologies are deemed to be important for future airports, but have yet to be widely implemented within present airports. While some insights into the adoption challenges were identified from the quantitative results, the differences in the extent of implementation among the various I4.0 technologies remain to be explored in future research in the context of an airport.

Organisational readiness has shown statistically significant influence on the successful adoption of I4.0 technologies in the airport. This is corroborated by Fig. 4 where top management support and an organisational culture that promotes digitalisation are one of the most frequently cited success factors by the airport operators surveyed. On the other hand, technological challenges such as technology complexity, technology compatibility and specific technology limitations were statistically insignificant in influencing successful I4.0 technology adoption in airports. This can possibly be attributed to the following reasons:

- Airport operators usually work with technology vendors in the development and the implementation of I4.0 technologies, and often these vendors provide free or highly affordable trials as a marketing tool to attract adoption (INT-9). Thus, technology complexities that involves extensive usability trials to address any operational concerns prior to technology adoption will not be a significant deterrent to airport operators in adopting I4.0 technologies.

- After a cost-benefit analysis, airport operators may recognize the added value of investing in I4.0 technologies despite the incompatibility with older systems. These I4.0 technologies can potentially improve operational efficiency, enhance passenger experience, as well as contribute to other business objectives for digital transformation (see **Table 9**), which may justify further investment in infrastructure to accommodate the new technologies.

- Airport operators may not be too concerned about specific technology limitations of I4.0 technologies as they do not necessarily have the technological expertise to accurately assess these technologies (INT-6). These limitations are likely to be of greater concern to technology vendors who may struggle to develop solutions that are suitable for the airport environment.

While technological challenges were most prevalently reported in literature when adopting I4.0 technologies in airports (Tan & Masood, 2021), the empirical results from the surveys and interviews have shown that most of the technological challenges do not have statistically significant influence on I4.0 technology adoption from the perspective of the airport operators. This is consistent with the observation made by Masood & Egger (2019) in a similar study relating to the adoption of AR where academic research tend to have a greater technological focus. An exception in this study is financial costs of the technologies which have shown to have statistically significant influence on the successful adoption of I4.0 technologies in airports.

Although passenger receptivity and acceptance was ranked fairly low in the list of success factors (Fig. 4), passenger receptivity has shown statistically significant influence on the successful adoption of I4.0 technologies in the airport. This is possibly because the involvement of passengers in pilot trials are important to assess the feasibility of the technological solutions. However, the eventual adoption of the technology could be the result of other objectives such as the mitigation of health and safety concerns, which may in some cases lead to inconvenience for passengers. Passenger receptivity is thus a measure of the success of technology adoption rather than having a direct influence on whether the technology is eventually adopted or not.

While stakeholder support did not have a statistically significant influence on the successful adoption of I4.0 technologies, stakeholder support is still crucial considering the multi-stakeholder environment within the airport. The reason for the statistical insignificance could possibly be due to the sample of survey respondents, where over 90% of the survey respondents work for private airport operators. According to qualitative inputs from the survey, non-privately-owned airports face greater pressure in garnering stakeholder support to get access to funds for digital transformation efforts. This financial pressure is likely to be less prominent for private airport operators.

Overall, the empirical results from the statistical tests provided generally good support for the theoretical model presented in Fig. 2.

## 9. Conclusions

The industry survey is the first-of-its-kind that was conducted to understand the implementation challenges that airport operators face in adopting I4.0 technologies in the airport. This has contributed towards addressing the research gap of the lack of empirical research into the challenges faced by airport operators in I4.0 technology adoption in airports.

The empirical study in this research has focused mainly on understanding the perspectives of airport operators, particularly the challenges they faced in technology adoption. As other airport stakeholders (ground handlers, airlines, concessionaires) are often the end-users of the technologies, this provides room for further research on how the extent of the challenges faced in I4.0 adoption differs among various stakeholders.

Furthermore, the survey results have shown that that the I4.0 technologies were not implemented to a similar extent in airports despite the generic challenges that were faced in adopting the various I4.0 technologies in the airport. This suggests that each I4.0 technology may have challenges or considerations that are specific to the technology or the airport. Thus, future research can look into studying the implementation of each I4.0 technology in greater detail in the context of their application within the airport environment. Also, as airports around the world vary in terms of their ownership structures (Neufville & Odoni, 2013), it will be interesting for future research to explore whether the technology adoption process and the challenges faced will differ based on the ownership structure of the airport. This will overcome any bias from the survey results which are predominantly completed by private operators of the airport.

As the world presses on in its fight against the COVID-19 pandemic, the airport industry needs to constantly adapt and adopt state-of-the-art technologies. By assimilating the challenges and success factors for adopting I4.0 technologies, airports will be better poised to address near-term issues relating to the pandemic and prepare for the post-pandemic surge in international travel.

**Acknowledgments:** This work was conducted as part of the *Industrial Systems of the Future* research program at the University of Cambridge, Cambridge, UK. The authors are most grateful to the participants of the industrial survey, reviewers and colleagues for their constructive and helpful comments. J.H. Tan acknowledges generous support of Changi Airport Group.

**Competing interests:** The authors have no competing interests.